\documentclass[trackchanges,twocolumn,times]{aastex701}

\begin{document}

\title{Can Little Red Dots Contribute To The Early Universe Cosmic Dust Budget?}

\author[0000-0002-5221-7557]{Chris~Ashall}
\email{chris.ashall24@gmail.com}
\affiliation{Institute for Astronomy, University of Hawai`i at M\={a}noa,
2680 Woodlawn Drive, Honolulu, HI 96822, USA}

\author[0000-0003-3997-5705]{Rohan P. Naidu}
\email{rnaidu@hawaii.edu}
\affiliation{Institute for Astronomy, University of Hawai`i at M\={a}noa,
2680 Woodlawn Drive, Honolulu, HI 96822, USA}

\author[0000-0001-5586-6950]{Alberto Torralba}
\email{Alberto.Torralba@ista.ac.at}
\affiliation{Institute of Science and Technology Austria (ISTA), Am Campus 1, 3400 Klosterneuburg, Austria}

\author[0009-0007-3791-7890]{Irene Shivaei}
\email{ishivaei@cab.inta-csic.es}
\affiliation{Centro deAstrobiología (CAB),CSIC-INTA,Cra. deAjalvirKm. 4, E-28850
Torrejón de Ardoz, Madrid, Spain}

\author[0000-0002-0302-2577]{John Chisholm}
\email{chisholm@austin.utexas.edu}
\affiliation{Department of Astronomy, The University of Texas at Austin,
Austin, TX 78712, USA}
\affiliation{Cosmic Frontier Center, The University of Texas at Austin,
Austin, TX 78712, USA}

\author[0000-0003-2488-4667]{Hanpu Liu}
\email{hanpu.liu@princeton.edu}
\affiliation{Department of Astrophysical Sciences, Princeton University,
4 Ivy Lane, Princeton, NJ 08544, USA}

\author[0009-0002-8965-1303]{Zhaoran Liu}
\email{zrliu@mit.edu}
\affiliation{MIT Kavli Institute for Astrophysics and Space Research,
70 Vassar Street, Cambridge, MA 02139, USA}

\author[0000-0003-2871-127X]{Jorryt Matthee}
\email{jorryt.matthee@ist.ac.at}
\affiliation{Institute of Science and Technology Austria (ISTA),
Am Campus 1, 3400 Klosterneuburg, Austria}

\author[0000-0001-7186-105X]{Kyle Medler}
\email{kmedler@hawaii.edu}
\affiliation{Institute for Astronomy, University of Hawai`i at M\={a}noa,
2680 Woodlawn Drive, Honolulu, HI 96822, USA}

\author[0000-0001-5888-2542]{Tyco Mera}
\email{tycomera@gmail.com}
\affiliation{Institute for Astronomy, University of Hawai`i at M\={a}noa,
2680 Woodlawn Drive, Honolulu, HI 96822, USA}

\author[0000-0003-1169-1954]{Takashi J. Moriya}
\email{takashi.moriya@nao.ac.jp}
\affiliation{National Astronomical Observatory of Japan,
National Institutes of Natural Sciences,
2-21-1 Osawa, Mitaka, Tokyo 181-8588, Japan}
\affiliation{Graduate Institute for Advanced Studies, SOKENDAI,
2-21-1 Osawa, Mitaka, Tokyo 181-8588, Japan}
\affiliation{School of Physics and Astronomy, Monash University,
Clayton, VIC 3800, Australia}

\author[0000-0003-1927-4397]{Devesh Nandal}
\email{devesh.nandal@cfa.harvard.edu}
\affiliation{Center for Astrophysics | Harvard \& Smithsonian,
60 Garden Street, Cambridge, MA 02138, USA}

\author[0000-0003-3769-9559]{Robert A. Simcoe}
\email{simcoe@space.mit.edu}
\affiliation{MIT Kavli Institute for Astrophysics and Space Research,
70 Vassar Street, Cambridge, MA 02139, USA}

\author[0009-0007-3791-7890]{Wendy Q. Sun}
\email{wendysun@mit.edu}
\affiliation{Institute for Astronomy, University of Hawai`i at M\={a}noa,
2680 Woodlawn Drive, Honolulu, HI 96822, USA}

\begin{abstract}
The origin of dust in the early Universe remains uncertain. We explore whether Little Red Dots (LRDs) could provide a high-redshift dust-production channel. Motivated by an analogy with Type IIn supernovae, we investigate whether LRDs may share the efficient dust-forming conditions. We consider dust formation in  outer winds and later in a shielded cold dense shell after the central source fades. Such dust may avoid a global remnant-phase reverse shock, while longer LRD lifetimes may promote grain growth. We model the SEDs of two low-redshift LRD analogs as a thermal pseudo-photosphere plus optically thin dust components, obtaining dust reservoirs of order $10^{2}$--$10^{3}\,M_\odot$. Although photometry cannot exclude pre-existing dust, the narrow Balmer components of both analogs are close to the Case B ratio and their narrow-line environments are metal-poor, disfavoring a dominant diffuse host-ISM origin for the inferred reservoir. Formation within the LRD outflow is consistent with our optical-depth, sublimation, and energy-balance checks. In this scenario, the large dust masses do not produce a strong optical attenuation because a clumpy or asymmetric distribution can leave the dominant optical sightlines relatively unobscured. Our population calculation shows that the LRD contribution to early-Universe dust production can range from negligible to dominant. At higher efficiencies, LRDs can approach or exceed the lower CCSN contribution above $z\approx5$, whereas less favorable assumptions yield a minor contribution. We therefore propose that LRDs can act as early-Universe dust factories and, under some conditions, may dominate over CCSNe or provide seed grains for subsequent growth in the early ISM. 

\end{abstract}

\keywords{\uat{Galaxies}{573} --- \uat{Cosmology}{343} --- \uat{Dust}{739} --- \uat{Supernovae}{847} }

\section{Introduction}
The rapid emergence of dust in the high-redshift Universe remains a major unresolved question in galaxy evolution  \citep[e.g.,][]{Witstok23,Markov24,Ormerod25,Nanni25, Shivaei25,Casey26}. Supernovae (SNe) have been proposed as dominant dust producers in the early Universe because their massive progenitors evolve on much shorter timescales than asymptotic giant branch stars \citep{TodiniFerrara01,Dwek07}. Nearby SNe and young SN remnants demonstrate that individual explosions can form substantial quantities of dust \citep{Matsuura11,Indebetouw14,DeLooze17,Medler25}. However, as the expanding ejecta collide with the surrounding circumstellar and interstellar material, a reverse shock propagates back through the ejecta, heating the gas and destroying grains through sputtering and grain--grain collisions. The fraction of the newly formed dust that survives this processing and is ultimately deposited into the interstellar medium therefore remains uncertain \citep{BianchiSchneider07,Nozawa07,Micelotta16}. Combined with the limited number of direct SN observations at z $\gtrsim$ 5 \citep{Coulter26}, this uncertainty leaves both the high-redshift SN rate and the effective dust-production efficiency of SNe poorly constrained, opening the possibility that other sources may contribute significantly to the early dust budget. 

With the launch of JWST, Little Red Dots (LRDs; \citealt{Matthee24}) have emerged as one of the most intriguing populations of high-redshift sources \citep[e.g.,][]{Harikane23, Greene24, Inayoshi25}. These compact, red sources are remarkably abundant across $z\approx2-10$ ($\approx10^{-5}$ cMpc$^{-3}$) -- one LRD is found on average in every moderately deep ($\sim25$ mag AB) NIRCam ($\approx10$ arcmin$^{-2}$) image \citep[e.g.,][]{ Kokorev24, Rinaldi26, Weibel26bhstars}. Initially thought to be massive galaxies \citep[e.g.,][]{Labbe23} or dusty AGN \citep[e.g.,][]{Kocevski23}, it has become clear that LRDs have a combination of properties that is unprecedented in other known types of sources. Briefly, unlike any known class of galaxies or AGN, LRDs are \textit{simultaneously} X-ray faint \citep[e.g.,][]{Yue24}, FIR/radio-quiet \citep[e.g.,][]{Akins24}, have unique V-shaped UV-optical SEDs \citep[e.g.,][]{Furtak24}, show strong Balmer breaks \citep[e.g.,][]{Wang24}, display deep Balmer absorption \citep[e.g.,][]{Torralba26}, and do not vary on year to decade timescales \citep[e.g.,][]{Liu26}.

Various lines of evidence suggest that the central engines of LRDs are a new type of astrophysical object -- ``Black Hole Stars" (BH*s) \citep[e.g.,][]{Naidu25BHstar, degraaff25bhstar, degraaff25pop, Sun26, Torralba26}. BH*s could be intermediate mass black holes (IMBHs) or supermassive stars (SMS) enshrouded in dense, hydrogen-rich gas envelopes \citep[e.g.,][]{Kido25, Begelman25, Inayoshi25, Liu25, Liu26synth, Sneppen26, Santarelli26, Roman-Garza26} such that they radiate in a manner reminiscent of black holes (e.g., broad lines, intense luminosity; \citealt{Matthee24, Labbe24monster, Torralba26gn9771}) as well as stellar phenomena (e.g., photospheric absorption features such as the Calcium triplet, water, and Balmer breaks; \citealt{Lin25Egg, Wang26, Torralba26}). A key insight from the BH* picture is that the redness of LRDs is set by gas physics, not dust. The red continuum reflects the $\approx5000$ K blackbody-like pseudo-photosphere \citep[e.g.,][]{degraaff25pop, Sun26, Umeda26}, and the extreme Balmer decrements of $\approx15$ \citep[e.g.,][]{Lin2026b} arise from radiative transfer processes in dense gas \citep[e.g.,][]{Chang26, Yan26, Nikopoulos25}. However, as we will discuss in this paper, it is plausible that LRDs show hints of \textit{freshly forming} dust, manifesting in the form of mid-infrared excesses above their pseudo-photospheric continuum.

Recently, \citet{Naidu26} drew a connection between LRDs and Type IIn SNe. The basic similarity is that both these classes of objects are likely powered by a central engine embedded within optically thick, hydrogen-rich material. In Type IIn SNe, narrow and intermediate-width H lines are produced when fast SN ejecta interact with dense, slowly moving,  H-rich CSM, forming an optically thick, partially ionized line-forming region \citep[e.g.,][]{Schlegel90,Dessart09,Smith17}. In LRDs, features mirroring IIn SNe may arise when a persistent fast wind or radiation from the central engine interacts with a slow, dense outer wind or envelope \citep[e.g.,][]{Matthee26, Naidu26,Sneppen26,Martins26, Chisholm2026}. As shown in \citet{Naidu26}, both LRDs and IIn SNe exhibit a pseudo-photospheric continuum, strong Balmer emission, P-Cygni-like absorption, and broad electron-scattering wings, suggesting that similar radiative-transfer processes operate in these dense, H-rich outflows (see \citealt{Naidu26} for a detailed list of parallels). The main differences are the total mass involved and the nature of the power source: LRDs may be sustained by a long-lived central engine, whereas SNe IIn are powered by a terminal stellar explosion. 

Given that Type IIn SNe are among the most efficient dust-forming transients observed locally \citep{Smith2009,Gall2014}, these similarities motivate the question of whether LRDs may also represent significant dust-production sites, particularly in the high-redshift Universe where LRDs are ubiquitous. We address this possibility in this work.


\section{Dust Formation in LRDs}
\label{sec:dust_formation}
Within the interacting-envelope framework, dust can form in two natural sites. The first is a cold dense shell (CDS) at the interface between the fast inner wind and slower outer wind; during the main LRD phase, it lies below the photosphere. Such shells efficiently form dust in interacting SNe because their post-shock gas can become dense, cool, and partially shielded, allowing molecules and grains to condense \citep[e.g.,][]{Smith08,Gall2014,Sarangi18,Shahbandeh25}. In SNe IIn, ongoing shock interaction can initially keep this gas too hot or strongly irradiated for grains to survive \citep[e.g.,][]{Sarangi18}. Likewise, a long-lived LRD engine may delay grain formation by maintaining the CDS above the condensation temperature. As the engine fades, the shell may cool below the characteristic $\sim$2000~K condensation temperature---whose precise value depends on grain composition and gas density---allowing rapid nucleation and growth \citep{Smith17}.

The second site is the extended cool wind above the photosphere, where the temperature falls with radius below the condensation temperature of refractory species. In the SN analogy, this resembles pre-existing circumstellar dust produced in dense progenitor outflows. Such dust is observed around mass-losing red supergiants, whose slow winds cool as they expand and allow molecules and oxygen-rich grains, including silicates and alumina, to condense \citep[e.g.,][]{Massey05,Verhoelst09,Cherchneff13,WalmswellEldridge12}. LBV-like eruptive mass loss provides a related analogy: dense expelled shells or nebulae can cool, become partially shielded, and form dust on larger scales \citep[e.g.,][]{Umana10,Agliozzo19,Agliozzo21}. Thus, in the LRD scenario, dust may form continuously in the cool, dense, slowly expanding envelope outside the pseudo-photosphere.

Molecule formation also provides an efficient cooling pathway in the dense envelopes and winds of massive stars and SNe, helping the gas reach conditions favorable for dust condensation \citep{Fassia01,Gerardy02,Medler25,Mera26,Mera26b}. Given the LRD--SN~IIn similarities, it may regulate the thermal structure of LRD envelopes and the conditions for grain growth; water absorption has already been detected in two LRDs \citep{Wang26}. Future observations should search for CO, SiO, and SiS \citep{Liljegren23,Rho24,Mera26}. Modeling these features can constrain the temperature, density, and chemical structure of the dust-forming regions, while resolved molecular emission could directly probe envelope asymmetries and clumping \citep{Mera26b}.

\begin{figure}
    \centering
    \includegraphics[width=\linewidth,  clip=True]{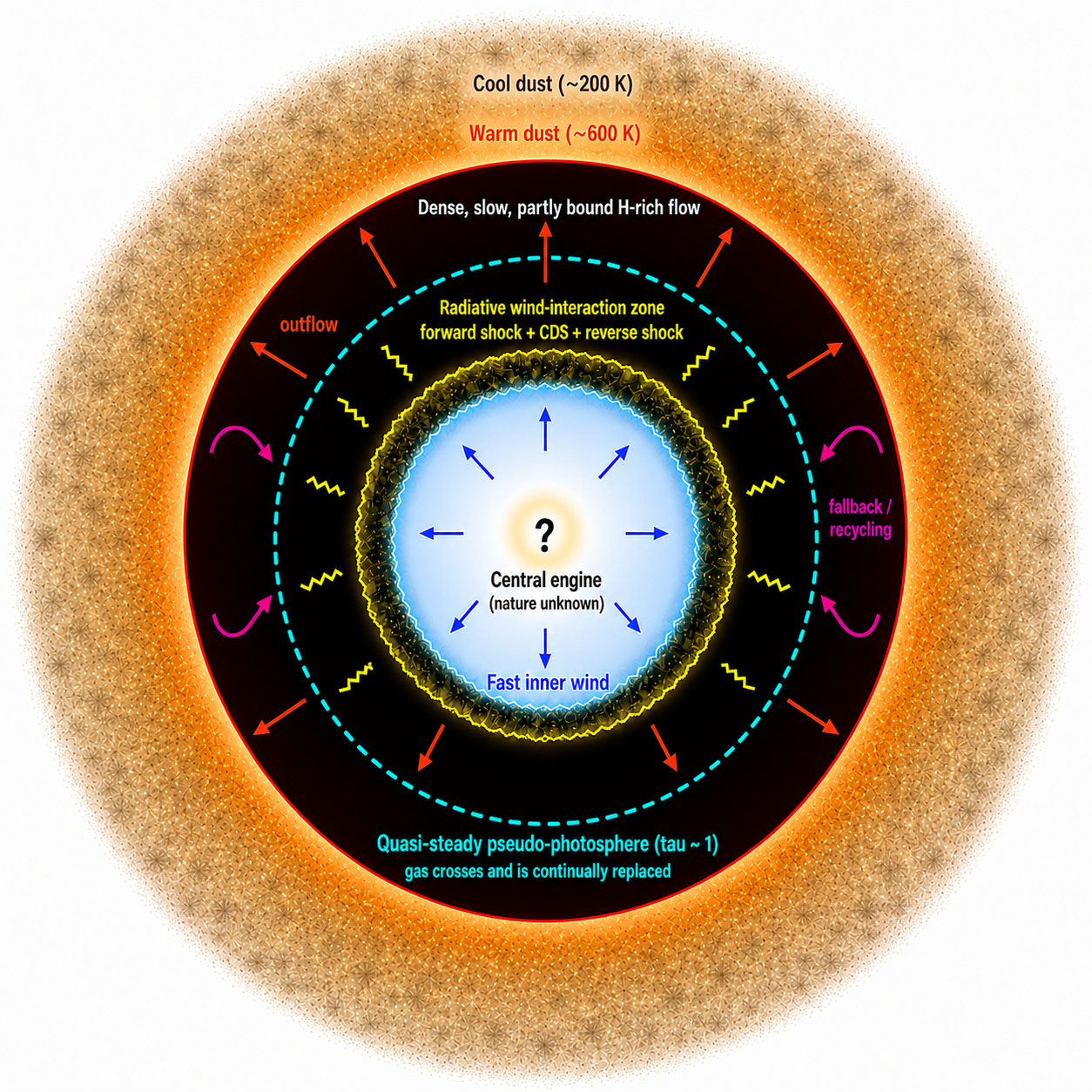}
    \caption{Schematic of the proposed emission and reprocessing regions in an LRD. A fast inner wind from an unspecified central engine interacts with a dense, slow, partly bound H-rich flow, continually regenerating radiative shocks and a cooled compressed layer. The surrounding optically thick gas absorbs, scatters, and reprocesses radiation from both the central engine and the wind interaction, producing a quasi-steady pseudo-photosphere at $\tau\sim$1. Gas continuously crosses this optical-depth surface and is replaced, so it does not represent a fixed material shell. Farther from the central source, the outflowing and recycled gas expands and cools; dense, shielded clumps may then provide favorable conditions for dust formation, transitioning from warm $\sim600$~K to cool $\sim200$ K dust. The schematic represents the flow during an active episode and is not to scale. }
    \label{fig:diagram}
\end{figure}

Figure~\ref{fig:diagram} shows a schematic one-dimensional representation of the regions where molecules and dust may form in LRDs. The overall structure is motivated by the wind-enshrouded central-engine picture of \citet{Naidu26}, while the dust-emitting regions are inspired by the Type IIn CDS framework of \citet{Shahbandeh25}. Maintaining a quasi-steady photospheric radius requires the envelope’s density and opacity structure to remain approximately stable. In this model, that stability is achieved by balancing the outward ram pressure of the fast inner wind against the pressure and inward ram pressure of the dense outer flow, it may even be the case that some of this outer material is moving inwards. Material launched below the fiducial escape velocity, \(v_{\rm esc}\sim500\,\mathrm{km\,s^{-1}}\), stalls and falls back toward the wind-interaction region, replenishing the envelope, preventing rapid density dilution, and sustaining the photosphere near a constant radius. In this picture, dust may form continuously within the extended outer wind, with additional delayed formation in the CDS after the central source fades. Unlike dust condensed within freely expanding SN ejecta, dust formed in an LRD wind or in the CDS would not necessarily encounter the global remnant-phase reverse shock produced when SN ejecta are decelerated by the surrounding medium. Such shocks propagate through the dust-bearing ejecta and can destroy newly formed grains through sputtering and grain--gas collisions. Because an LRD is instead sustained by a long-lived central engine and continuous outflow, dust in the outer wind or a radiatively cooled shell may avoid this system-wide processing. Moreover, LRD lifetimes of potentially $\gtrsim10~{\rm Myr}$ \citep[e.g.,][]{Sun26, Santarelli26, Roman-Garza26} may provide substantially more time for dust to form, accumulate, and grow within dense, shielded regions before being injected into the surrounding medium. LRDs could therefore produce substantially larger cumulative dust yields per source than individual SNe.

The model shown in Figure \ref{fig:diagram} is a one dimensional schematic diagram. In reality, the geometry of the system may be highly aspherical \citep[e.g.,][]{Matthee26, Madau26, deugenio26}. Such asymmetries are observed in SNe IIn and could produce dense, shielded clumps in the wind or CDS, creating localized regions where molecules and dust can form more efficiently \citep[e.g.,][]{Hoffman08, Priestley19,Mera26b, Tang26, Ji26}. Furthermore, it is not yet clear whether the central source is truly continuous or instead undergoes episodic phases of enhanced activity, analogous to repeated mass ejections in pulsational pair-instability events \citep[e.g.,][]{Woosley07,Woosley17}. If the power source turns on and off, successive episodes could build up concentric shells of material, providing multiple dense interfaces where molecules and dust may form -- see \citealt{Nandal26winds} for examples of such episodic winds in supermassive stars which may be progenitors of BH*s \citep[e.g.,][]{Zwick26, Martins26, Chisholm2026, Naidu26}.

\section{Abundances and refractory elements}

Dust forms through the condensation of refractory elements, including C, O, Mg, Si, Al, and Fe, with the resulting grain composition set by the elemental abundances and physical conditions within the outflow. The spectra of LRDs contain numerous metal lines -- most prominently Fe and O -- demonstrating that chemically enriched material is present \citep[e.g.,][]{Lin25Egg, Tripodi25,  Torralba26gn9771, deugenio25, PerezGonzalez26}. 
These observations show that LRDs contain at least some of the elements required for dust formation, although the total refractory-element mass and the dominant grain composition remain uncertain. 

However, in theory, the final dust yield depends on the available refractory-element mass and also the efficiency with which it condenses into grains. In H- and He-rich massive-star winds, including LBV eruptions, refractory elements represent only a small fraction of the total expelled mass. Observed dust-to-ejected-mass ratios are typically a few $\times10^{-3}$ to $\sim10^{-2}$, corresponding to approximately $0.2$--$1\%$ of the expelled material \citep{Gomez2010,Agliozzo14,Agliozzo21}. Core-collapse SN ejecta contain larger masses of newly synthesized refractory elements and can produce $\sim0.1$--$1~M_\odot$ of dust, equivalent to approximately $1$--$10\%$ of the ejecta mass for a typical massive-star explosion \citep{Priestley2020,Niculescu-Duvaz22}. 

The origin of the metal reservoir available for dust formation remains an important uncertainty in our work, and future work focusing on the evolutionary pathway to LRDs will help determine the resulting dust composition.
If LRDs are formed out of pre-existing accreting BHs, the refractory elements must primarily be inherited from gas previously enriched by star formation in the host galaxy or surrounding protocluster environment \citep{Mayer2015,ChonOmukai2020}. If the LRD phase traces the formation or evolution of a supermassive star \citep[e.g.,][]{NandalLoeb2026, Zwick26, Martins26}, nuclear processing could provide an additional local source of enriched material provided that processed layers are expelled through winds or pulsational mass loss, or that the star undergoes an explosive rather than direct-collapse fate \citep{Inayoshi2013,Gieles2018,Nagele2023}. Further, in the dense globular cluster-like environment where the SMS is expected to form  \citep[e.g.,][]{Chisholm2026}, early SNe in the immediate vicinity may also enrich the extended dense gas envelope \citep{Gieles2018,Jimenez2021}.

Regardless of the central engine, the observed metal lines, outflow signatures, and Type~IIn-like spectral morphology imply that enriched CSM-like material is already present. Dust formation during the observed LRD phase therefore does not require contemporaneous metal production. Instead, the engine may primarily maintain a dense pseudo-photosphere, drive winds and shocks, and create the cool, shielded regions in which the pre-existing enriched material can condense into dust.

\section{Dust Fitting}
\label{sec:dust_fitting}

Motivated by the physical similarities between SNe IIn and LRDs, we apply dust-fitting methods commonly used for SNe \citep{Shahbandeh23} to two low-redshift LRDs from \citet{Park26}, selected for their infrared coverage and high signal-to-noise data (see also \citealt{Lin2026b}). We model each SED as a pseudo-photospheric blackbody plus two optically thin dust components, as is typical in SN dust modeling \citep{Fox2010,DerKacy26}. We consider a hot graphite component paired either with cooler astronomical silicate (C+Si) or cooler graphite (C+C).

We use this deliberately simple decomposition as a proof of concept and to obtain first-order estimates of the dust masses and compositions. For optically thin dust, the emergent flux scales as
\begin{equation}
F_\nu \propto
\frac{M_{\rm d}\,\kappa_\nu(a)\,B_\nu(T_{\rm d})}{D_L^2},
\end{equation}
\noindent where $M_{\rm d}$ is the dust mass, $T_{\rm d}$ is the dust temperature, $\kappa_\nu(a)$ is the grain-size-dependent mass absorption opacity, $B_\nu(T_{\rm d})$ is the Planck function, and $D_L$ is the luminosity distance. We compute luminosity distances from the source redshifts assuming a flat cosmology with $H_0 = 70~{\rm km~s^{-1}~Mpc^{-1}}$ and $\Omega_m = 0.3$, and adopt wavelength-dependent graphite and astronomical silicate opacities from \citet{DraineLi07}. The fitted parameters are the temperatures and masses of the graphite and silicate dust components, together with the temperature and normalization of the photospheric blackbody. We fix the hot graphite grain radius to $a=0.1~\mu{\rm m}$ and allow the cool-component grain radius to vary. With only a few photometric points in the MIR, it is difficult to accurately constrain the grain sizes. Unlike in SNe, the potential absence of a reverse shock in these objects may allow small grains to survive. On the other hand, the long lifetimes of LRDs may provide sufficient time for grains to grow to larger sizes. Detailed models of dust formation and grain growth under these conditions will be needed to determine which of these effects dominates. Furthermore, MIRI/MRS spectra of these or similar objects would allow the dust composition and grain properties to be constrained more accurately.

We note that the blackbody radii reported below are luminosity-equivalent emitting radii derived from the fitted dust luminosities and temperatures; they do not represent physical dust locations or shell radii. The minimum shell radii required for optically thin emission are calculated separately
in Section~\ref{sec:physical-consistency}. We also include a free--free component, motivated by its importance in interacting systems with dense circumsource material \citep{DerKacy26,Dessart26phot}. Rather than fitting it independently, we fix its amplitude to $9\%$ of the near-infrared continuum, corresponding to $F_\nu(3~\mu{\rm m})=0.00581~{\rm mJy}$ for J0129+0628 and $F_\nu(3~\mu{\rm m})=0.0354~{\rm mJy}$ for J0829+1312, and adopt $F_\nu\propto\nu^{-0.1}$ and $T_e=10^4~{\rm K}$.

\begin{figure}
    \centering
    \includegraphics[width=\linewidth,  clip=True]{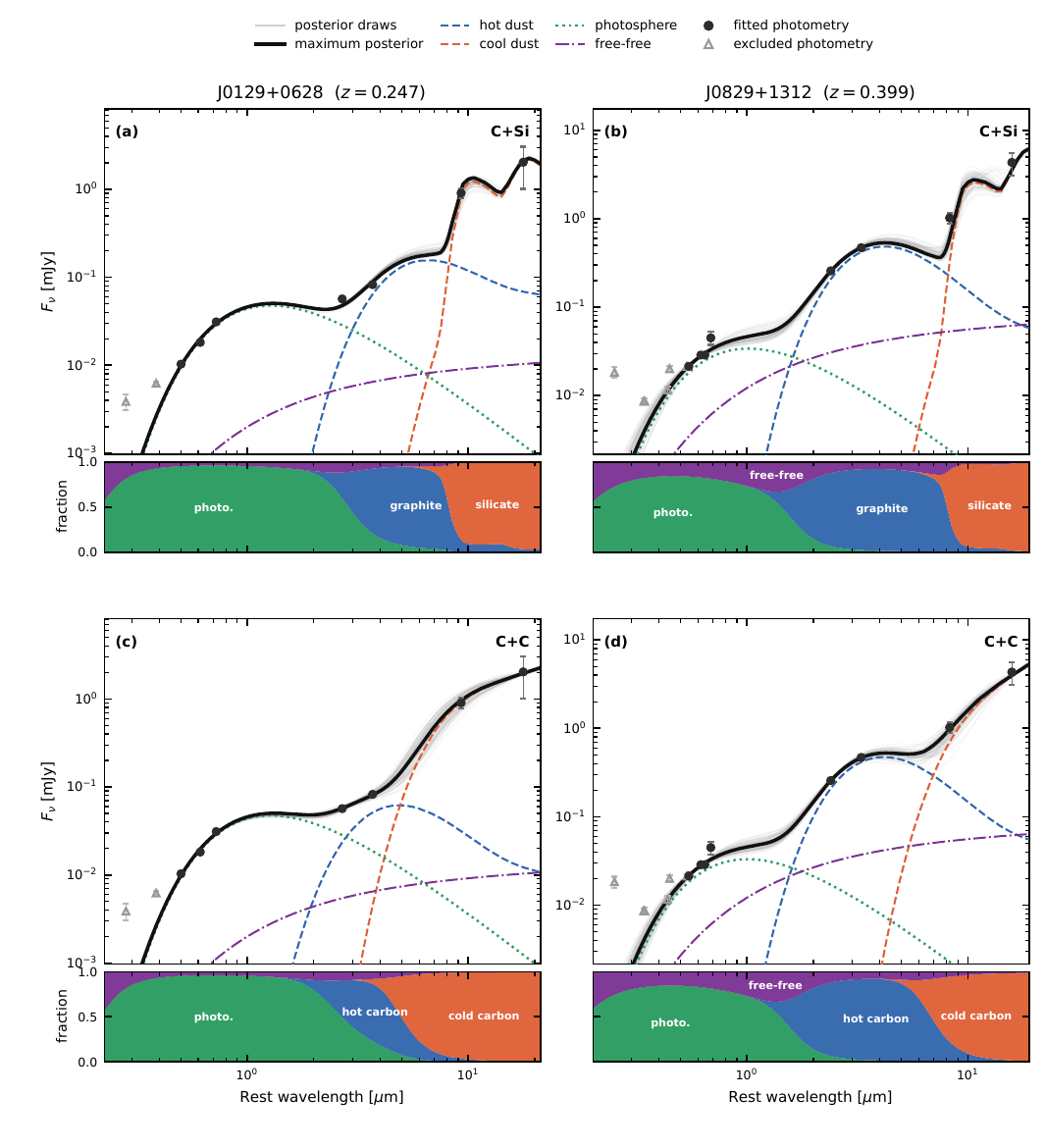}
    \caption{Rest-frame SEDs of the two low-redshift LRD analogs from
\citet{Park26}, J0129+0628 (left) and J0829+1312 (right). The upper row shows the hot-graphite plus cool-silicate (C+Si) model, while the lower row shows the  hot- plus cool-graphite (C+C) model. In both cases, the broadband SEDs are reproduced by a $\sim4000$--$5000$ K pseudo-photosphere, two optically thin mid-infrared dust components, and a fixed free--free contribution. Black curves show the maximum-posterior models, thin gray curves show posterior draws, filled circles indicate fitted photometry, and open triangles indicate excluded measurements. The lower panels show the fractional contribution of each model component. Both composition families reproduce the available photometry,
showing that the broadband data do not uniquely determine the dust composition. Both adopted model families nevertheless require a substantial infrared-emitting dust reservoir.}
    \label{fig:dustmodel2}
\end{figure}

To estimate the formal statistical uncertainties within each adopted model family, we sampled the posterior distributions using an affine-invariant Markov chain Monte Carlo method. We used 48 walkers for 2000 steps and adopted uniform priors for the temperatures. We sampled the dust masses, pseudo-photospheric radius, and fitted cool-component grain radius with uniform priors in logarithmic space over the ranges listed in Table~\ref{tab:sed-results}. The black curves in Figure~\ref{fig:dustmodel2}
represent the maximum-posterior models, whereas Table~\ref{tab:sed-results}
reports posterior medians with uncertainties defined by the 16th and 84th
percentiles. We note that these intervals quantify statistical uncertainty conditional on the adopted model and do not include the larger systematic uncertainties associated with the opacity, geometry, or additional emission components.


\begin{deluxetable*}{lcccccc}
\tablecaption{SED Model Assumptions, Priors, and Posterior Constraints\label{tab:sed-results}}
\tabletypesize{\tiny}
\tablewidth{0pt}
\tablehead{
\colhead{Parameter} &
\colhead{C+Si prior/fixed} &
\colhead{C+C prior/fixed} &
\colhead{J0129+0628 C+Si} &
\colhead{J0829+1312 C+Si} &
\colhead{J0129+0628 C+C} &
\colhead{J0829+1312 C+C}
}
\startdata
$T_{\rm phot}$ (K) & $[4000,6000]$  & $[4000,6000]$  &
$4015^{+24}_{-11}$ & $5044^{+366}_{-337}$ &
$4042^{+53}_{-31}$ & $5041^{+381}_{-339}$ \\
$R_{\rm phot}$ ($10^{16}$ cm) & $[10^{-3},10^{4}]$  & $[10^{-3},10^{4}]$  &
$1.200^{+0.014}_{-0.020}$ & $1.179^{+0.219}_{-0.177}$ &
$1.173^{+0.027}_{-0.045}$ & $1.181^{+0.225}_{-0.184}$ \\
$T_{\rm hot}$ (K) & $[100,2500]$  & $[350,2500]$  &
$470^{+5}_{-8}$ & $715^{+13}_{-17}$ &
$631^{+46}_{-42}$ & $727^{+25}_{-22}$ \\
$M_{\rm d,hot}$ ($M_\odot$) & $[1,10^{10}]$  & $[10^{-3},10^{10}]$  &
$1.09^{+0.17}_{-0.06}$ & $1.23^{+0.24}_{-0.16}$ &
$0.104^{+0.069}_{-0.042}$ & $1.09^{+0.30}_{-0.25}$ \\
$R_{\rm BB,d}^{\rm hot}$ ($10^{16}$ cm) & Derived & Derived &
$64.7^{+3.9}_{-1.7}$ & $97.0^{+6.9}_{-4.8}$ &
$25.2^{+5.5}_{-4.6}$ & $92.7^{+8.9}_{-8.6}$ \\
$T_{\rm cool}$ (K) & $[20,1200]$ & $[20,350]$  &
$189^{+5}_{-8}$ & $179^{+12}_{-12}$ &
$246^{+15}_{-16}$ & $191^{+14}_{-15}$ \\
$M_{\rm d,cool}$ ($M_\odot$) & $[10^{2},10^{10}]$  & $[10^{-3},10^{10}]$  &
$114^{+22}_{-10}$ & $(1.07^{+0.45}_{-0.31})\times10^{3}$ &
$175^{+37}_{-28}$ & $(2.71^{+0.96}_{-0.63})\times10^{3}$ \\
$a_{\rm cool}$ ($\micron$) & $[10^{-3},10]$  & $[10^{-3},10]$  &
$0.086^{+0.422}_{-0.079}$ & $0.042^{+0.605}_{-0.038}$ &
$0.014^{+0.085}_{-0.012}$ & $0.047^{+0.205}_{-0.043}$ \\
$R_{\rm BB,d}^{\rm cool}$ ($10^{16}$ cm) & Derived & Derived &
$729^{+48}_{-27}$ & $(2.15^{+0.30}_{-0.24})\times10^{3}$ &
$559^{+64}_{-45}$ & $(2.23^{+0.41}_{-0.30})\times10^{3}$ \\
$M_{\rm d,tot}$ ($M_\odot$) & Derived & Derived &
$115^{+22}_{-10}$ & $(1.07^{+0.45}_{-0.31})\times10^{3}$ &
$175^{+37}_{-28}$ & $(2.72^{+0.96}_{-0.63})\times10^{3}$
\enddata
\end{deluxetable*}

\subsection{Two-component dust models}
Below we discuss the C+C and C+Si fits. In both cases, the warmer component dominates the near- to mid-infrared emission, whereas a substantially more massive cold component accounts for the longer-wavelength flux.

We treat these fits as illustrative, phenomenological decompositions rather than unique solutions. The available broadband photometry provides limited leverage on the detailed dust temperature distribution, grain composition, optical depth, geometry, and possible additional emission components. Systematic uncertainties, particularly those associated with the adopted opacity and grain properties, therefore dominate over the formal statistical intervals. We emphasize order-of-magnitude dust masses and use the C+C and C+Si models to bracket composition-dependent solutions.

\subsubsection{C+C model}

The C+C fits are shown in Figure~\ref{fig:dustmodel2}. For J0129+0628, the warm graphite component has $T_{\rm d}=631^{+46}_{-42}~{\rm K}$, $M_{\rm d}=0.104^{+0.069}_{-0.042}~M_\odot$, and $R_{\rm BB,d}=2.52^{+0.55}_{-0.46}\times10^{17}~{\rm cm}$. The cold graphite component has $T_{\rm d}=246^{+15}_{-16}~{\rm K}$, $M_{\rm d}=175^{+37}_{-28}~M_\odot$, and $R_{\rm BB,d}=5.59^{+0.64}_{-0.45}\times10^{18}~{\rm cm}$. The fitted photosphere has $T_{\rm phot}=4042^{+53}_{-31}~{\rm K}$ and $R_{\rm phot}=1.173^{+0.027}_{-0.045}\times10^{16}~{\rm cm}$.

For J0829+1312, the warm graphite component has $T_{\rm d}=727^{+25}_{-22}~{\rm K}$, $M_{\rm d}=1.09^{+0.30}_{-0.25}~M_\odot$, and $R_{\rm BB,d}=9.27^{+0.89}_{-0.86}\times10^{17}~{\rm cm}$. The cold component has $T_{\rm d}=191^{+14}_{-15}~{\rm K}$, $M_{\rm d}=2.71^{+0.96}_{-0.63}\times10^3~M_\odot$, and $R_{\rm BB,d}=2.23^{+0.41}_{-0.30}\times10^{19}~{\rm cm}$. The corresponding photospheric parameters are $T_{\rm phot}=5041^{+381}_{-339}~{\rm K}$ and $R_{\rm phot}=1.181^{+0.225}_{-0.184}\times10^{16}~{\rm cm}$.

Using the posterior median values, the warm-to-cold dust-mass ratios are approximately $5.9\times10^{-4}$ and $4.0\times10^{-4}$ for J0129+0628 and J0829+1312, respectively. These small ratios primarily reflect the much larger mass required for low-temperature graphite to reproduce the observed long-wavelength luminosity.

The C+C fits therefore demonstrate that the infrared excess can be reproduced using carbonaceous grains alone. However, the large inferred cold-dust masses are sensitive to the adopted graphite opacity and motivate consideration of an alternative composition for the long-wavelength component.

\subsubsection{C+Si model}

The C+Si fits are shown in Figure~\ref{fig:dustmodel2}. In this model, the warm component remains graphite, while the cold component is represented by astronomical silicate. The small free--free contribution shown in the figure remains fixed rather than being fitted independently.

For J0129+0628, the graphite component has $T_{\rm d}=470^{+5}_{-8}~{\rm K}$, $M_{\rm d}=1.09^{+0.17}_{-0.06}~M_\odot$, and $R_{\rm BB,d}=6.47^{+0.39}_{-0.17}\times10^{17}~{\rm cm}$. The silicate component has $T_{\rm d}=189^{+5}_{-8}~{\rm K}$, $M_{\rm d}=114^{+22}_{-10}~M_\odot$, and $R_{\rm BB,d}=7.29^{+0.48}_{-0.27}\times10^{18}~{\rm cm}$. The photospheric component has $T_{\rm phot}=4015^{+24}_{-11}~{\rm K}$ and $R_{\rm phot}=1.200^{+0.014}_{-0.020}\times10^{16}~{\rm cm}$. 
The C+Si model provides an alternative description of the
$5$--$20\,\micron$ emission of J0129+0628, showing that the broadband SED is
compatible with a cool silicate component. However, it is not statistically
preferred over the C+C model. Spectroscopic detection of the 9.7 or
$18\,\micron$ silicate features will be required to establish the presence
and strength of silicate emission.

For J0829+1312, the graphite component has $T_{\rm d}=715^{+13}_{-17}~{\rm K}$, $M_{\rm d}=1.23^{+0.24}_{-0.16}~M_\odot$, and $R_{\rm BB,d}=9.70^{+0.69}_{-0.48}\times10^{17}~{\rm cm}$. The silicate component has $T_{\rm d}=179^{+12}_{-12}~{\rm K}$, $M_{\rm d}=1.07^{+0.45}_{-0.31}\times10^3~M_\odot$, and $R_{\rm BB,d}=2.15^{+0.30}_{-0.24}\times10^{19}~{\rm cm}$. The corresponding photospheric parameters are $T_{\rm phot}=5044^{+366}_{-337}~{\rm K}$ and $R_{\rm phot}=1.179^{+0.219}_{-0.177}\times10^{16}~{\rm cm}$.

Using the posterior median values, the graphite-to-silicate mass ratios are approximately $9.6\times10^{-3}$ for J0129+0628 and $1.1\times10^{-3}$ for J0829+1312.

Replacing cold graphite with astronomical silicate reduces the inferred cold-dust mass by factors of approximately $1.5$ and $2.5$ for J0129+0628 and J0829+1312, respectively. This difference demonstrates the strong systematic dependence of the inferred masses on the adopted grain composition and wavelength-dependent opacity. We therefore interpret the fitted mass ratios as composition-dependent estimates of the emitting dust, not direct measurements of the relative carbon- and silicate-forming elemental abundances.

\subsection{Origin of the inferred dust reservoir}
\label{sec:dust_origin_caveat}

We interpret the fitted masses as the amount of dust participating in the observed infrared emission under the assumptions of our SED model. Broadband photometry does not uniquely determine when that dust formed: the reservoir may contain material present before the observable LRD phase, dust formed or grown during the LRD phase in the extended wind or dense shielded regions, or a mixture of both.

The same distinction applies to the Type IIn analogy, where the dense CSM may contain both pre-existing and newly formed dust. We therefore report the fitted masses as model-dependent infrared-emitting reservoirs and treat the newly formed, surviving yield as a separate quantity that we vary in the population calculation.

The pre-LRD state of these systems is unknown, and the evolutionary
path into the LRD phase is especially uncertain at high redshift. We therefore
cannot exclude pre-existing circumnuclear dust heated, exposed, or redistributed
by the central source. For the two analogs considered here, however, the
component-separated line luminosities reported by \citet{Lin2026b} give narrow-component
$\mathrm{H}\alpha/\mathrm{H}\beta\simeq3.31$ and $3.17$ for J0129+0628 and
J0829+1312, respectively. These ratios are close to the Case B value and
indicate little attenuation in the narrow-line-emitting gas. Both objects also
belong to the metal-poor DESI LRD population. We therefore disfavor a dominant
diffuse host-ISM origin for the inferred dust reservoir.

\subsubsection{Physical consistency of the optically thin wind model}
\label{sec:physical-consistency}

{

We next assess whether dust formed within the proposed wind can satisfy
constraints from optical depth, sublimation, and energy balance. The schematic
in Section~2 describes a long-lived, stratified outflow rather than a detached
dust shell, with grains heated by radiation, shocks, or grain--gas collisions.
We therefore separate direct constraints on the emitting material from
quantities that depend on its heating history and geometry.

We distinguish an emitting area from a physical dust location. The radii
reported in Table~\ref{tab:sed-results} are
luminosity-equivalent radii,
\begin{equation}
R_{\rm BB,d}=\left(\frac{L_{\rm d}}
{4\pi\sigma T_{\rm d}^{4}}\right)^{1/2},
\end{equation}
and describe an equivalent emitting area, not the distance of the grains from
the source or the boundary of a shell.  Consequently,
$R_{\rm BB,d}<R_{\rm phot}$ does not place the dust beneath the
pseudo-photosphere.

Following the thin-shell approximation used for SN dust
\citep{Shahbandeh23,Shahbandeh25}, the optical depth of dust distributed over
a fraction $f_\Omega$ of a spherical shell is
\begin{equation}
\tau_\lambda \simeq
\frac{M_{\rm d}\kappa_\lambda}
{4\pi f_\Omega R_{\mathrm{sh}}^{2}}.
\end{equation}
We use the fitted masses and opacities and impose the conservative requirement
$\tau_{\rm peak}\leq0.1$ at the peak of each component. This gives
$R_{\rm sh}\gtrsim4.3\times10^{17}$--$1.7\times10^{18}$ cm for the warm
components and $R_{\rm sh}\gtrsim9.1\times10^{18}$--$4.2\times10^{19}$ cm
for the cool components when $f_\Omega=1$.  All exceed
$R_{\rm phot}\simeq1.2\times10^{16}$ cm and correspond to flow times of
$3\times10^{2}$--$3\times10^{4}$ yr at $500\,\mathrm{km\,s^{-1}}$, well
below the adopted LRD lifetime.  The bounds scale as $f_\Omega^{-1/2}$.  A
clumpy wind is compatible with the calculation when its emitting
structures are also optically thin in the MIR, although they may remain
opaque at UV and optical wavelengths (see Section~5).

The fitted temperatures, $T_{\rm d}=179$--727 K, are below commonly adopted sublimation temperatures for silicate and carbonaceous grains. Both grain populations are therefore thermally stable under the conditions required by the SED fits. Their survival is consistent with formation or residence in the cooler, shielded regions of the stratified wind, outside the hottest shocked gas, further supporting the physical plausibility of our wind interpretation.

In our long-lived wind model, radiative and mechanical power can supply the dust luminosity. Radiation from the central source and shocks below the pseudo-photosphere is thermalized and re-emerges through the wind, while shocks at larger radii can heat grains through radiation and collisions. Continuous formation can also maintain a population of newly formed grains as they cool. The observed \(L_{\rm phot}\) therefore need not be the sole source of \(L_{\rm dust}\), and broadband photometry does not determine the relative contributions of these heating channels.

Together, these checks show that the fitted SED components admit a physically consistent realization within the proposed long-lived wind. The dust can reside outside the pseudo-photosphere on scales reachable within the LRD lifetime, remain optically thin to its own MIR emission, survive below its sublimation temperature, and draw on radiative and mechanical energy in the wind. The broadband data do not fix the geometry or heating history, but they are consistent with the optically thin, stratified outflow shown in Section~2.
}

\section{Dust emission without strong observed optical obscuration}

In our interpretation, dust emission and the optical extinction inferred from the observed spectrum are not necessarily tightly coupled. The MIR/FIR-emitting dust must absorb a fraction of the UV--optical luminosity and reradiate this energy thermally. However, a large total dust mass does not automatically imply strong extinction along the lines of sight that dominate the observed optical emission.

If the dust has a clumpy, equatorial, porous shell-like, or otherwise asymmetric distribution, as is often observed in interacting SNe, different regions of the source may experience substantially different levels of attenuation. The observed optical emission will then be luminosity-weighted toward the least-obscured regions, which can outshine more heavily attenuated parts of the system. In the SN~IIn-like scenario considered here, the optical continuum emerges from the surface of the optically thick H-rich pseudo-photosphere, while the Balmer emission is powered by shock interaction and forms in the dense shocked and photoionized circumstellar material at or above the photosphere. These less-obscured regions may dominate the observed optical spectrum even if dust elsewhere absorbs and reradiates a significant fraction of the UV--optical luminosity in the MIR/FIR.

The optically thin infrared emission does not imply negligible UV--optical attenuation; it only indicates that the dust is largely transparent to its own MIR/FIR thermal emission. A substantial dust reservoir can therefore be distributed over a large volume or confined to dense clumps with an incomplete covering factor. Some regions may be strongly attenuated and power much of the infrared luminosity, while the observed optical spectrum remains dominated by less-obscured photospheric and interaction-region emission. If part of the MIR emission arises from newly formed dust that later cools, this picture also predicts MIR-bright LRDs without a substantial cold-dust component, as possibly reported in \citet{Delvecchio25}.

\section{Population Dust-Production Rates}
\label{sec:rates}

If the large dust masses inferred for individual LRDs are representative, we
ask whether LRDs can make a significant contribution to the cosmic dust
budget at high redshift. The relevant quantity is the
comoving dust-production-rate density, $\dot{\rho}_{\rm dust}(z)$, which
depends on the abundance of sources, the dust mass produced per source, and
the timescale over which that dust is assembled.

Because the LRD dust yield, event multiplicity, survival fraction,
and fraction of dust-forming sources remain uncertain, we treat this
calculation as an illustrative parameter study rather than a unique
prediction. The left panel of Figure~\ref{fig:lrd_vs_sn} presents a useful
reference comparison, while the right panel shows how its normalization
changes under alternative population assumptions.

\subsection{LRD Contribution to the Cosmic Dust Budget}

We estimate the LRD contribution using the comoving number density of LRDs,
$n_{\rm LRD}(z)$, from \citet{Kapoor26}, who fit the observed LRD abundance
across $0\lesssim z\lesssim10$ with a log-normal function of the form
\begin{equation}
n_{\rm LRD}(z)=\phi_0\,f(z)\,
\exp\!\left[-\frac{\left(\ln(1+z)-\mu_z\right)^2}
{2\sigma_z^2}\right],
\label{eq:lrdlognormal}
\end{equation}
where $f(z)=(1+z)^{3/2}/[s\,(1+z)^{1/2}-1]^2$ accounts for the
cosmological volume element, with
$\phi_0=3\times10^{-6}\,{\rm cMpc^{-3}}$,
$\mu_z=\ln(1+z_0)$, $z_0=5$, $\sigma_z=0.28$, and $s=0.903$.
This abundance peaks at $z\approx5$ with a value of order
$3\times10^{-5}\,{\rm cMpc^{-3}}$ and declines toward both lower and higher
redshift; extrapolated regions beyond the range of the input data are shown
explicitly in Figure~\ref{fig:lrd_vs_sn}. This fit is calibrated to LRDs with
rest-frame optical magnitudes $M_{5100}\approx-21$, integrated down to
$M_{5100}\approx-20$, and we adopt a per-object dust yield consistent with
the same luminosity range.

We interpret $n_{\rm LRD}(z)$ as the number density of active LRD
host galaxies and adopt a total active lifetime of
$\tau_{\rm LRD}=10~{\rm Myr}$ \citep[e.g.,][]{Sun26}. Clustered star
formation provides a physical motivation for multiple central-engine
episodes within this interval \citep[e.g.,][]{Belokurov23,Fujimoto24}. For
our reference case, we adopt $N_{\rm events,eff}=30$, motivated by the
globular-cluster statistics discussed by \citet{Naidu26}.
Here, $N_{\rm events,eff}$ counts only independent episodes that process
fresh material; repeated processing of the same gas or dust reservoir is
represented by a lower effective event number.

The surviving dust mass associated with each LRD event is parameterized as
\begin{equation}
M_{\rm d,LRD}=\varepsilon_{\rm L}\,M_{\rm out}\,f_{\rm surv,LRD},
\label{eq:mdlrd}
\end{equation}
where $M_{\rm out}$ is the dust-forming gas reservoir,
$\varepsilon_{\rm L}$ is the effective dust-conversion efficiency, and
$f_{\rm surv,LRD}$ is the surviving dust fraction. The luminosities of the
optical-to-near-infrared blackbody components inferred for LRDs imply central
black-hole masses of order $10^{4}$--$10^{6}\,M_\odot$, for example if the
observed luminosity is emitted near the Eddington limit
\citep[e.g.,][]{Lin2026b,Umeda26,Greene25}.

For the reference calculation, we adopt a characteristic
dust-forming outflow reservoir of $M_{\rm out}=10^{5}\,M_\odot$
\citep{Chen26,Naidu26}. We treat this value as an uncertain normalization
rather than a measurement for every LRD; the resulting rates scale linearly
with $M_{\rm out}$.

We also adopt $f_{\rm surv,LRD}=0.8$, representing a scenario in which
dust formed in a persistent LRD outflow avoids the global remnant-phase
reverse-shock destruction experienced by SN ejecta. We motivate the explored
range of $\varepsilon_{\rm L}$ through comparisons with dust formation in
LBVs and interaction-dominated SNe
\citep{Agliozzo14,Agliozzo21,Niculescu-Duvaz22}. For LRDs, we write the
effective fraction of outflow mass converted into dust schematically as
\begin{equation}
\varepsilon_{\rm L}\simeq
X_{\rm ref,LRD}f_{\rm cond,LRD},
\label{eq:epsilon-lrd}
\end{equation}
where $X_{\rm ref,LRD}$ is the refractory-element mass fraction of the
dust-forming material and $f_{\rm cond,LRD}$ is the fraction of those elements
that condenses into grains. Thus, $\varepsilon_{\rm L}$ incorporates both the
available refractory-element budget and the condensation efficiency and is
the quantity used in Equations~\ref{eq:mdlrd} and \ref{eq:rholrd}. If LRDs and
interaction-dominated SNe share similar condensation physics, their
dust-conversion efficiencies should show the same first-order dependence on
the available refractory material.

We therefore consider
$\varepsilon_{\rm L}=0.1\%,0.5\%,1\%,5\%,$ and $10\%$. The lower values
allow for ordinary enriched outflows, lower refractory abundances, or
incomplete condensation, whereas the 5--10\% values represent limiting
processed- or ejecta-like outflows motivated by the SN analogy. For the
reference values of $M_{\rm out}$ and $f_{\rm surv,LRD}$, these efficiencies
correspond to surviving dust masses of 80, 400, 800, 4000, and
8000\,$M_\odot$ per event, respectively. These values are also consistent
with ALMA constraints of less than $10^{6}\,M_\odot$ of colder dust in LRDs
at wavelengths longer than those studied here \citep{Casey25}. This range
covers the composition of the dust-forming material rather than the mean
LRD metallicity. High-redshift SNe can form dust from newly
synthesized metals, and SMS or BH-star interpretations of LRDs similarly
leave open the possibility of processed outflow material.

The resulting LRD dust-production-rate density is
\begin{equation}
\dot{\rho}_{\rm dust,LRD}(z)
=n_{\rm LRD}(z)\,f_{\rm dusty}
\frac{N_{\rm events,eff}\,\varepsilon_{\rm L}\,M_{\rm out}\,
f_{\rm surv,LRD}}{\tau_{\rm LRD}},
\label{eq:rholrd}
\end{equation}
where $f_{\rm dusty}$ is the fraction of LRDs that experience the
required dust-forming conditions. In the left panel of
Figure~\ref{fig:lrd_vs_sn}, we set $N_{\rm events,eff}=30$,
$f_{\rm surv,LRD}=0.8$, and $f_{\rm dusty}=1$. In the right panel, we show
that all LRD curves can be rescaled without changing their adopted redshift
dependence according to
\begin{equation}
\mathcal{S}=
\left(\frac{N_{\rm events,eff}}{30}\right)
\left(\frac{f_{\rm surv,LRD}}{0.8}\right)f_{\rm dusty}.
\label{eq:lrdscaling}
\end{equation}

The nature and evolution of the central engine may also affect the LRD dust
yield. If LRDs are powered by an accreting black hole, X-ray
and UV emission from the accretion flow could heat, sublimate, or destroy dust
in the inner regions. Efficient dust survival may therefore require formation
at sufficiently large or shielded radii, or after the accretion luminosity
has declined.

\subsection{CCSNe Contribution to the Cosmic Dust Budget}

We compare this with the dust-production-rate density from core-collapse SNe.
We adopt the cosmic star-formation-rate density, $\psi(z)$, from
\citet{MadauDickinson14} at $z\lesssim9$ and the high-redshift \textit{JWST}
constraints from \citet{Weibel26} at $z\gtrsim9$. For a \citet{Chabrier03}
IMF, stars with initial masses $8\leq M/M_\odot\leq50$ give one CCSN per
$82.5\,M_\odot$ of star formation, so that
\begin{equation}
R_{\rm CC}(z)=\frac{\psi(z)}{82.5\,M_\odot}.
\label{eq:rcc}
\end{equation}
The surviving dust mass per CCSN is
\begin{equation}
M_{\rm d,SN}=\varepsilon_{\rm SN}\,\langle M_{\rm ej}\rangle\,
f_{\rm surv,SN},
\label{eq:mdsn}
\end{equation}
where $\varepsilon_{\rm SN}$ is the corresponding effective dust-conversion efficiency,
$\langle M_{\rm ej}\rangle$ is the mean ejecta mass, and
$f_{\rm surv,SN}$ is the reverse-shock survival fraction. For the reference
CCSN model we adopt $\langle M_{\rm ej}\rangle=10.9\,M_\odot$,
$\varepsilon_{\rm SN}=5\%$, and $f_{\rm surv,SN}=5\%$, giving
$M_{\rm d,SN}=0.027\,M_\odot$ of surviving dust per event. To bracket the
large uncertainty in the effective surviving yield, in
Figure~\ref{fig:lrd_vs_sn} we show a band spanning
$\varepsilon_{\rm SN}=0.1\%$--$10\%$
\citep[e.g.,][]{BianchiSchneider07,Nozawa07,Gall11,Sarangi18,Marassi19}.
The CCSN dust-production-rate density is then
\begin{equation}
\dot{\rho}_{\rm dust,SN}(z)
=R_{\rm CC}(z)M_{\rm d,SN}
=\frac{\psi(z)}{82.5\,M_\odot}\,
\varepsilon_{\rm SN}\,\langle M_{\rm ej}\rangle\,f_{\rm surv,SN}.
\label{eq:rhosn}
\end{equation}

We note that the predicted CCSN rates are sensitive to the adopted IMF and
the progenitor mass range assumed to produce core-collapse events, and that
our calculation neglects contributions from PISNe. Nevertheless, this
comparison remains informative, and our predicted rates are broadly
consistent with the observed CCSN rates recently measured by the JADES
Transient Survey out to $z\sim5$ \citep{DeCoursey26}.

\subsection{LRDs versus CCSNe}

\begin{figure*}
    \centering
    \includegraphics[width=\linewidth,clip]{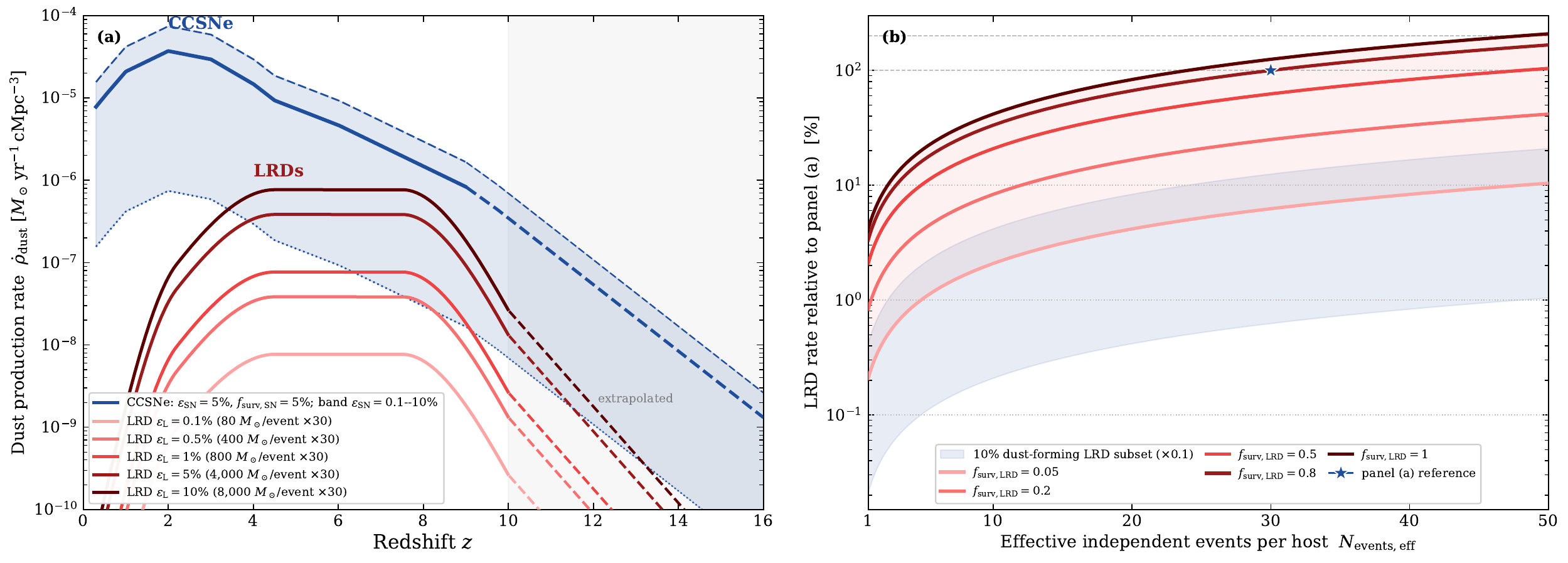}
    \caption{Left: Comoving dust-production-rate density,
    $\dot{\rho}_{\rm dust}(z)$, for CCSNe and LRDs. The CCSN rate follows the
    cosmic star-formation-rate density and is shown for a central model with
    $\varepsilon_{\rm SN}=5\%$ and $f_{\rm surv,SN}=5\%$; the shaded region
    spans $\varepsilon_{\rm SN}=0.1\%$--$10\%$. The LRD rate follows the
    log-normal number-density fit of \citet{Kapoor26}, which peaks near
    $z\approx5$, and is shown for
    $\varepsilon_{\rm L}=0.1\%,0.5\%,1\%,5\%,$ and $10\%$, assuming
    $M_{\rm out}=10^{5}\,M_\odot$,
    $N_{\rm events,eff}=30$, $\tau_{\rm LRD}=10~{\rm Myr}$,
    $f_{\rm surv,LRD}=0.8$, and $f_{\rm dusty}=1$. The adopted LRD abundance
    is luminosity limited, and lower-luminosity LRDs may alter this relation.
    Solid curves indicate the redshift range covered by the adopted input
    data, while dashed curves show extrapolated regions. Right:
    Rescaling of the LRD dust-production rate with the effective
    number of independent events and dust-survival fraction. The star marks
    the assumptions adopted in the left panel, while the blue region shows
    the corresponding range if only $10\%$ of LRDs form dust.}
    \label{fig:lrd_vs_sn}
\end{figure*}

In the left panel of Figure~\ref{fig:lrd_vs_sn}, we compare the LRD
and CCSN dust-production-rate densities. The CCSN
contribution follows the cosmic star-formation history, whereas the LRD curves
follow the adopted LRD number density and peak near $z\approx5$. At this
redshift, the $\varepsilon_{\rm L}=5\%$ and $10\%$ reference models remain
approximately factors of 20 and 10 below the central CCSN prediction,
respectively, but exceed the lower edge of the CCSN band by factors of
approximately 2.6 and 5.2. These cases identify the parameter range in which
LRDs could provide a non-negligible dust-production channel.

In the right panel, we vary the uncertain population parameters.
Lower survival fractions, fewer independent events, or
a dust-forming subset of the LRD population reduce the inferred rate
linearly; for example, if only $10\%$ of LRDs form dust, all LRD rates
decrease by one order of magnitude. Conversely, larger event multiplicities
or survival fractions increase the normalization. The reference comparison
can therefore be rescaled directly as these quantities become better
constrained.

The broad CCSN band reflects the substantial uncertainties in both the amount
of dust condensed in each explosion and the fraction that survives subsequent
destruction by the reverse shock. Consequently, the relative LRD and CCSN
contributions depend on both the LRD population parameters and whether the
substantial dust masses observed in nearby SNe and remnants ultimately
survive and reach the ISM.

Recent work has suggested that LRDs may be linked to globular
clusters in formation, with their compact sizes, high stellar densities, and
number densities potentially connecting them to the progenitors of
present-day metal-poor globular clusters
\citep[e.g.,][]{Rantala26,Williams26,Chisholm2026}. This provides a physical
motivation for our exploration of multiple events per host, but does not determine
their number or whether they process independent reservoirs; these
uncertainties are represented by $N_{\rm events,eff}$ in the right panel. In
this possible picture, the rest-frame UV emission may arise from a compact
young stellar population, while the red optical continuum is powered by a
central SMS or IMBH embedded in an optically thick envelope. The associated
wind could provide dense, shielded regions for molecule and dust formation
and could subsequently expel material into cooler outer regions.

Building on the source-level picture above, we find that LRDs could
provide an important contribution under the processed-outflow, high-survival,
and repeated-event scenarios represented by the upper curves, whereas less
favorable assumptions yield a minor contribution. The calculation therefore
maps how the LRD contribution changes across the allowed population
parameters. Larger spectroscopic samples and direct far-infrared constraints
will be needed to determine which regime is representative. Our comparison
includes only LRDs and CCSNe; future work should also incorporate AGB stars,
grain growth in the ISM, and subsequent dust destruction.

\section{Conclusions}

The remarkable spectral similarities between LRDs and SNe IIn suggest that
both may be shaped by the same broad physical configuration: a central power
source embedded within a massive, optically thick, slowly moving envelope,
into which a faster wind or outflow deposits energy. The main differences are
the nature and scale of the power source. SNe IIn are produced by terminal
stellar explosions interacting with dense H-rich CSM, whereas LRDs may be
powered by long-lived central engines embedded in much more massive gas
reservoirs. We use the analogy with a scaled-up, persistent version of
the interaction-dominated phase of SNe IIn as our working physical framework
and adopt an active lifetime of order $\sim10~{\rm Myr}$.

SNe IIn are among the most efficient dust-forming transients observed locally.
Dust may be present in their progenitor winds before explosion, and additional
dust can form in the dense, shielded CDS produced by ejecta--CSM interaction.
LRDs may provide analogous conditions, but on larger mass and time scales. In
this picture, dust may form continuously in the extended outer wind, analogous
to dust formation in the dense outflows of evolved massive stars or during
LBV-like eruptions, and later in the CDS after the central engine turns off and
the weakening radiation field allows the gas to cool below the
dust-condensation temperature. Thus, although some dust may form in
shielded outer regions while the central engine is active, a substantial
fraction of the eventual newly condensed dust may be produced after the
central luminosity declines and a larger fraction of the outflow can cool
below the condensation temperature. Because LRDs are powered by persistent
engines rather than terminal explosions, dust in their outflows would not
necessarily undergo the same late reverse-shock processing that affects SN
ejecta, although shocks and sputtering within the LRD wind could still destroy
grains. Their long lifetimes may further allow grains to grow to larger sizes
in dense, shielded regions.

We demonstrate that the SEDs of two low-redshift LRD analogs can be reproduced
by a thermal pseudo-photosphere together with optically thin
graphite+silicate or graphite+graphite dust components, extending the analogy
with SNe IIn into the infrared. The inferred total emitting dust
masses span approximately $10^{2}$--$10^{3}\,M_\odot$, depending on
the adopted dust composition and model assumptions. These masses describe
the infrared-emitting dust reservoirs in these two analogs. Our
physical-consistency checks show that dust formed within the LRD outflow
admits a viable realization in the proposed wind model: it can reside outside
the pseudo-photosphere on scales reachable within the adopted LRD lifetime,
remain optically thin to its own MIR emission, survive thermal sublimation,
and be powered by radiative and mechanical energy in the wind.

Broadband photometry cannot determine whether the emitting dust formed
during the LRD phase, was already present in the host or circumnuclear
environment, or is a mixture of both, nor can it determine what fraction will
ultimately survive. However, the near-Case-B narrow Balmer ratios of both
analogs, together with their membership in the metal-poor DESI LRD population,
disfavor a dominant contribution from the diffuse host-galaxy ISM.
Pre-existing circumnuclear dust remains possible, but these results motivate
our proposal that dust formation within the LRD outflow may supply a
substantial fraction of the inferred reservoir.

We also examine whether LRDs could make a significant contribution to the
cosmic dust budget at high redshift. To capture the population-level
uncertainty, we vary the dust-conversion efficiency, effective number of
independent events, survival fraction, and fraction of dust-forming LRDs.
The inferred LRD contribution follows the observed LRD abundance
and peaks near $z\approx5$. For the higher-efficiency reference cases,
$\varepsilon_{\rm L}\gtrsim5\%$, the LRD contribution can exceed the lower
edge of the CCSN uncertainty band over much of $z\approx4$--10, although it
remains below the central CCSN prediction. Under these assumptions, LRDs could
provide an important additional dust-production channel. Conversely,
lower dust-conversion efficiencies, fewer independent events, lower survival
fractions, or dust formation in only a small subset of LRDs result in a minor
or negligible contribution. If reverse-shock destruction substantially
reduces the surviving SN dust yield, LRDs could become more competitive with
the CCSN contribution. In this regime, LRD outflows may provide seed grains
that subsequently grow in the ISM, particularly at early epochs where there
has been less time for efficient grain growth.

CCSNe have traditionally provided the primary framework for rapid dust
enrichment in the early Universe because their massive progenitors evolve on
short timescales. However, no confirmed CCSN has yet been observed at
$z>5$, and their event rates, dust yields, and reverse-shock survival
fractions in this regime are therefore not directly measured. By contrast,
LRDs are observed in substantial numbers at $z\gtrsim5$, and the low-redshift
analogs modeled here show infrared excesses consistent with substantial dust
reservoirs. While determining the exact nature of the LRD central engine is
beyond the scope of this work, the radiative-transfer physics, line formation,
and dense wind structure may be similar to those in SNe IIn. We
therefore identify LRDs as a plausible high-redshift pathway for dust
production whose prevalence and net yield remain to be measured.

Future observations will determine which of these regimes applies.
\textit{JWST}/NIRSpec and MIRI/MRS spectroscopy of low-redshift analogs will
be required to constrain dust composition, heating, and
geometry beyond broadband photometry. Larger samples and rest-frame
mid-infrared and far-infrared constraints will test whether comparable dust
reservoirs are common among high-redshift LRDs. Improved LRD number densities,
duty cycles, and time-domain measurements of CCSN rates from \textit{JWST}
will then clarify whether LRDs are an unimportant or significant contributor
to the rapid dust enrichment observed at early times.

\begin{acknowledgments}
 We thank Kevin Park for sharing his data on J0129+0628 and J0829+1312.  C.A. acknowledges support from NASA grants 
JWST-GO-03726, JWST-GO-05057, JWST-GO-05290, JWST-GO-06023,
JWST-GO-06677,  JWST-GO-06583. Support for
programs \#3726, \#5057, \#5290,
\#6023, \#6677, and \#6583, 
were provided by NASA through a grant from the Space Telescope Science
Institute, which is operated by the Association of Universities for Research in
Astronomy, Inc., under NASA contract NAS 5-03127.
\end{acknowledgments}

\bibliography{sample701}{}
\bibliographystyle{aasjournalv7}

\end{document}